\documentclass[onecolumn, floatfix, preprintnumbers, eqsecnum, letterpaper, superscriptaddress, nofootinbib]{revtex4}
\usepackage{graphicx}
\usepackage{microtype}
\usepackage{amsmath}
\usepackage{amssymb}
\usepackage{subfigure}
\usepackage{hyperref}
\usepackage{url}
\usepackage{xcolor}
\usepackage{color}
\usepackage{mathrsfs}
\usepackage{calrsfs}
\usepackage{amsfonts}
\usepackage{eufrak}
\usepackage{tabularx}
\usepackage{eucal}
\usepackage{latexsym}
\usepackage{ragged2e}
\usepackage{epsfig}
\usepackage{textcomp}

\usepackage{caption}
\usepackage{subfigure}

\DeclareCaptionJustification{justified}{\leftskip=0pt \rightskip=0pt \parfillskip=0pt plus 1fil}
\definecolor{vividviolet}{rgb}{0.62, 0.0, 1.0}
\definecolor{amaranth}{rgb}{0.9, 0.17, 0.31}
\definecolor{palatinateblue}{rgb}{0.15, 0.23, 0.89}
\definecolor{brightpink}{rgb}{1.0, 0.0, 0.5}
\definecolor{cornflowerblue}{rgb}{0.39, 0.58, 0.93}
\definecolor{deepcarminepink}{rgb}{0.94, 0.19, 0.22}
\definecolor{radicalred}{rgb}{1.0, 0.21, 0.37}

\hypersetup{ linktoc=all,
    colorlinks, linkcolor={palatinateblue},
    citecolor={brightpink}, urlcolor={amaranth}
}

\graphicspath{{Images/}}

\graphicspath{{Images/}}

\def\@fnsymbol#1{\ensuremath{\ifcase#1\or \ddagger \or  $\textleaf$  \or \dagger
\else\@ctrerr\fi}}%

\makeatother
\def\sideremark#1{\ifvmode\leavevmode\fi\vadjust{\vbox to0pt{\vss% the remark
 \hbox to 0pt{\hskip\hsize\hskip1em%                          will appear only
 \vbox{\hsize1.3cm\tiny\raggedright\pretolerance10000%          on the side
 \noindent #1\hfill}\hss}\vbox to8pt{\vfil}\vss}}}%

\def\beq{\begin{equation}}
\def\eeq{\end{equation}}

\newcommand{\od}{\mathrm{d}}

\begin{document}

\title{Effective Pressure of the FRW Universe}

\author{Shi-Bei Kong}
\email{shibeikong@ecut.edu.cn}
\affiliation{School of Science, East China University of Technology, Nanchang 330013, China}

\begin{abstract}

In this paper, we study the effective pressure of the $N$-dimensional FRW(Friedmann-Robertson-Walker) universe in Einstein gravity, Gauss-Bonnet gravity, and Lovelock gravity. The effective pressure is defined by $P_{eff}:=-\od E/\od V$, where $E=\rho V$ is the effective energy and
$V$ is the volume of the FRW universe inside the apparent horizon. The effective pressure in Einstein gravity is always negative and its absolute value decreases with the horizon radius $R_A$. The effective pressure in Gauss-Bonnet gravity is different with the one in Einstein gravity only when $N\geq6$.
In this case, if $\alpha>0$, the effective pressure is always negative, but if $\alpha<0$, it is not always negative and has a minimum. The effective pressure in Lovelock gravity can have multiple zero-points and extreme points. The effective pressure in different dimensions has interesting relations. We also find that under certain condition, the effective pressure is equivalent with the `ordinary' pressure $p$ of the perfect fluid, and this condition do not depend on the specific choice of gravitational theories.

\end{abstract}

\maketitle

\section{Introduction}

In previous studies\cite{Kong:2021dqd,Abdusattar:2021wfv,Kong:2022xny,Abdusattar:2023hlj}, we investigated the thermodynamics,
equation of state and phase transitions for the FRW(Friedmann-Robertson-Walker) universe in various theories of gravity.
In all these cases, we found that the first law of thermodynamics for the FRW universe can be written as
\begin{alignat}{1}
\od E=-T\od S+W\od V. \label{fl}
\end{alignat}
It looks like a total differential, but is actually not\footnote{I thank Yi-Hao Yin to raise this question.}
because both the entropy $S$ and the thermodynamic volume $V$ are functions of the apparent horizon radius $R_A$.
From the thermodynamic perspective, it means that heat and work are not independent,
which is different from a usual thermodynamic system. What's more, the following relations are not meaningful
\begin{alignat}{1}
T=-\left(\frac{\partial E}{\partial S}\right)_{V},  \quad W=\left(\frac{\partial E}{\partial V}\right)_{S}.
\end{alignat}
Therefore, strictly speaking, ($T, S$) and ($W, V$) are not conjugate thermodynamic variables,
which also obscures their thermodynamic meanings. There is also a minus sign before $T\od S$,
which is not commonly seen for a usual thermodynamic system except de Sitter spacetime\footnote{De Sitter spacetime can be regarded 
as a special case of the FRW universe.}
and has been debated for decades\cite{Banihashemi:2022htw}.

Recently, we find a natural way to solve the above questions at the same time. The two terms in the first law of thermodynamics
(\ref{fl}) should be combined, i.e. it should be written as
\begin{alignat}{1}
\od E=-P_{eff}\od V, \label{first}
\end{alignat}
or
\begin{alignat}{1}
\od E=T_{eff}\od S, \label{second}
\end{alignat}
where $P_{eff}$ or $T_{eff}$ can be called the effective pressure or effective temperature of the FRW universe respectively,
thermodynamic volume $V$ or entropy $S$ is still a function of the apparent horizon radius $R_A$.
Obviously, $(P_{eff}, V)$ or $(T_{eff}, S)$ are conjugate pairs, and there is no minus sign problem before $T_{eff}\od S$.
The effective pressure or effective temperature measures the change of the energy with the change of the volume or the entropy, thus called `effective'.

In this paper, we concentrate on the first form (\ref{first}), and study the effective pressure $P_{eff}$.\footnote{The effective temperature
is investigated in another paper\cite{Kong:2025abc}.} The expression of the effective pressure can be obtained from (\ref{first})
\begin{alignat}{1}
P_{eff}=-\frac{\od E}{\od V},
\end{alignat}
which is usually a function of the apparent horizon radius $R_A$. At first, we get the effective pressure of the $N$-dimensional FRW universe in Einstein gravity, which is always negative and increases with $R_A$ and $N$. We also write the effective energy in a `Smarr' form and get its enthalpy and free energy.
Gauss-Bonnet gravity is an important theory of modified gravity, and it has been shown the Gauss-Bonnet term must appear in lower energy limit of the heterotic superstring $E_8\times E_8$\cite{Zwiebach:1985uq}. It is also found that in Gauss-Bonnet gravity, many things are affected by the 
spacetime dimension, such as in the gravitational collapse\cite{Brassel:2024erf} and Joule-Thomson expansion\cite{Abdusattar:2024sgk}. Therefore, we expect that the Gauss-Bonnet coupling $\alpha$ and spacetime dimension $N$ play roles in the effective pressure of the FRW universe, and our result shows that it is different with the one from Einstein gravity in $N\geq 6$. We also find that if $\alpha>0$, the effective pressure is always negative, but if $\alpha<0$, 
the effective pressure is positive when $R_A<\sqrt{-(N-4)(N-5)\alpha}$ or negative when $R_A>\sqrt{-(N-4)(N-5)\alpha}$ and has a minimum at $R_A=\sqrt{-2(N-4)(N-5)\alpha}$ with the minimum effective pressure
\begin{alignat}{1}
P_{min}=\frac{(N-2)(N-3)}{8\kappa_N(N-4)(N-5)\alpha}.
\end{alignat}
The $P_{eff}-R_{A}$ curve looks very similar with
the potential-radius curve between molecules, and more interestingly, the minimum effective pressure of the $N$-dimensional FRW universe lies
exactly on the $P_{eff}-R_{A}$ curve of the $N$-dimensional FRW universe. Lovelock gravity\cite{Lovelock:1971yv,Lovelock:1972vz} 
is a natural extension of Einstein gravity and takes Gauss-Bonnet gravity as its special case. It has also been used in the study of the FRW universe, 
such as its solution in the Lovelock gravity\cite{Nikolaev:2024gqo}. In this paper, we also consider Lovelock gravity and study the effective pressure of the FRW universe in this gravity. We find that the effective pressure of the FRW universe in Lovelock gravity may have multiple zero-points and extreme points. 
We also make a comparison of the effective pressure $P_{eff}$ with the pressure $p$ of the perfect fluid, and find that if $\dot{R}_A=HR_{A}$, they are equal regardless of the choice of the gravitational theories.

This paper is organized in the following way. In Sec.II, we make a brief introduction of the $N$-dimensional FRW universe.
In Sec.III, we study the effective pressure of the $N$-dimensional FRW universe in Einstein gravity.
In Sec.IV, we study the effective pressure of the $N$-dimensional FRW universe in Gauss-Bonnet gravity.
In Sec.V, we study the effective pressure of the $N$-dimensional FRW universe in Lovelock gravity. In Sec. VI, we conclude this paper
and make some discussion.

In this paper, we use natural units that $c=\hbar=1$.

\section{A Brief Introduction of the $N$-dimensional FRW Universe}

The metric for the $N$-dimensional FRW universe can be written as\cite{Cai:2006rs}
\begin{alignat}{1}
\od s^2=-\od t^2+a^2(t)\left(\frac{\od r^2}{1-kr^2}+r^2\od\Omega_{N-2}^2\right), \label{metric}
\end{alignat}
where $a(t)$ is the time-dependent scale factor\footnote{In the following, we use $a$ instead of $a(t)$ for convenience.}, $k$ is the spatial curvature (with $+1,0,-1$ corresponding to spatial closed, flat and open respectively), $\od\Omega^2_{N-2}$ 
is the metric of the co-dimension 2 unit sphere.
One can also  rewrite the metric as
\begin{alignat}{1}
\od s^2=h_{ab}\od x^a\od x^b+R^2\od\Omega_{N-2}^2,
\end{alignat}
where $R:=a(t)r$ is the physical radius and $h_{ab}$ is the $2$-dimensional metric with $a,b=0,1$ and $x^0=t, x^1=r$.

The apparent horizon of the FRW universe is defined by
\begin{alignat}{1}
h^{ab}\nabla_a R\nabla_b R=0,
\end{alignat}
and its solution is
\begin{alignat}{1}
R_A=\frac{1}{\sqrt{H^2+\frac{k}{a^2}}}. \label{AH}
\end{alignat}
A very useful relation for the apparent horizon is
\begin{alignat}{1}
\dot{R}_A=-HR_A^3\left(\dot{H}-\frac{k}{a^2}\right), \label{dAH}
\end{alignat}
where $``\cdot"$ means $\od/\od t$.

The FRW universe is filled with a perfect fluid, which energy-momentum tensor is written as
\begin{alignat}{1}
T_{\mu\nu}=(\rho+p)U_{\mu}U_{\nu}+pg_{\mu\nu}, \label{pf}
\end{alignat}
where $\rho$ is its energy density, $p$ is its pressure and $U^{\mu}$ is its 4-velocity.
The energy-momentum tensor satisfies the conservation condition $\nabla_{\mu}T^{\mu\nu}=0$,
and its `0' component is the continuity equation
\begin{alignat}{1}
\dot{\rho}+(N-1)H(\rho+p)=0. \label{continue}
\end{alignat}

Up to now, no specific theory of gravity is used, but the expression of the effective pressure is depended on the specific theory.
In the following investigations, we use three kinds of gravitational theories, i.e. Einstein gravity, Gauss-Bonnet gravity and Lovelock gravity.

\section{Effective Pressure of the FRW Universe in Einstein Gravity}

In Einstein gravity, the $N$-dimensional field equations can be written as\cite{Maharaj:2021gua}
\begin{alignat}{1}
R_{\mu\nu}-\frac{1}{2}g_{\mu\nu}R=\kappa_N T_{\mu\nu}, \label{fe}
\end{alignat}
where $\kappa_N$ is the gravitational coupling constant in $N$-dimension.
Insert the metric (\ref{metric}) of the FRW universe and the energy-momentum tensor (\ref{pf})
into the Einstein field equation (\ref{fe}), one can get the Friedmann's equation
\begin{alignat}{1}
H^2+\frac{k}{a^2}=\frac{2\kappa_N}{(N-1)(N-2)}\rho,
\end{alignat}
which provides us the expression of the energy density.
From the continuity equation (\ref{continue}), one can get the expression of the pressure
\begin{alignat}{1}
p=-\frac{N-2}{\kappa_N}\left(\dot{H}-\frac{k}{a^2}\right)-\frac{(N-1)(N-2)}{2\kappa_N}\left(H^2+\frac{k}{a^2}\right).
\end{alignat}
Using (\ref{AH}) and (\ref{dAH}), one can rewrite $\rho$ and $p$ as:
\begin{alignat}{1}
\rho=\frac{(N-1)(N-2)}{2\kappa_N R_A^2}, \quad p=\frac{(N-2)\dot{R}_A}{\kappa_N HR^3_A}-\frac{(N-1)(N-2)}{2\kappa_N R^2_A}. \label{ed}
\end{alignat}
We also give the work density\cite{Cai:2006rs,Hayward:1997jp} for comparison
\begin{alignat}{1}
W:=-\frac{1}{2}h^{ab}T_{ab}=\frac{1}{2}(\rho-p)=\frac{(N-1)(N-2)}{2\kappa_N R^2_A}-\frac{(N-2)\dot{R}_A}{2\kappa_N HR^3_A}.
\end{alignat}

From the energy density (\ref{ed}), one can get the effective energy of the FRW universe
\begin{alignat}{1}
E=\rho V=\frac{(N-2)}{2\kappa_N}\mathcal{A}_{N-2}R_A^{N-3}, \label{energy}
\end{alignat}
where $V=\mathcal{A}_{N-2} R_A^{N-1}/(N-1)$\cite{Cai:2006rs} is the volume of the $N$-dimensional FRW universe inside the apparent horizon and $\mathcal{A}_{N-2}=2\pi^{(N-1)/2}/\Gamma((N-1)/2)$\cite{Maharaj:2021gua} is the area of $(N-2)$-dimensional unit sphere.
As discussed in\cite{Maharaj:2021gua}, it should be noted that only in 4-dimensional spacetime, $\kappa_N/(2\mathcal{A}_{N-2})$ is equal to one.
One can see that (\ref{energy}) is just the Misner-Sharp energy inside the apparent horizon.
The effective energy of the FRW universe can be regarded as a function of $V$, so one can write the first law
of thermodynamics as\footnote{It has no ``heat" term, i.e. we treat the effective entropy of the FRW universe as a constant $S_{eff}=S_0$.}
\begin{alignat}{1}
\od E=-P_{eff}\od V.
\end{alignat}
Here $P_{eff}$ is the effective pressure
\begin{alignat}{1}
P_{eff}=-\frac{\od E}{\od V}=-\left(\frac{\od E}{\od R_A}\right)/\left(\frac{\od V}{\od R_A}\right)=-\frac{(N-2)(N-3)}{2\kappa_N R_A^2},
\end{alignat}
which vanishes for $N=2,3$ and its absolute value decreases with $R_A$. This effective pressure is equivalent to the `ordinary' pressure $p$ when $\dot{R}_A=HR_A$, and equivalent to $-W$ when $\dot{R}_{A}=2HR_A$, see Appendix for short proofs. The effective pressure for $N=4,5,6$ is shown below.

\begin{figure}[h]
\includegraphics[height=5cm]{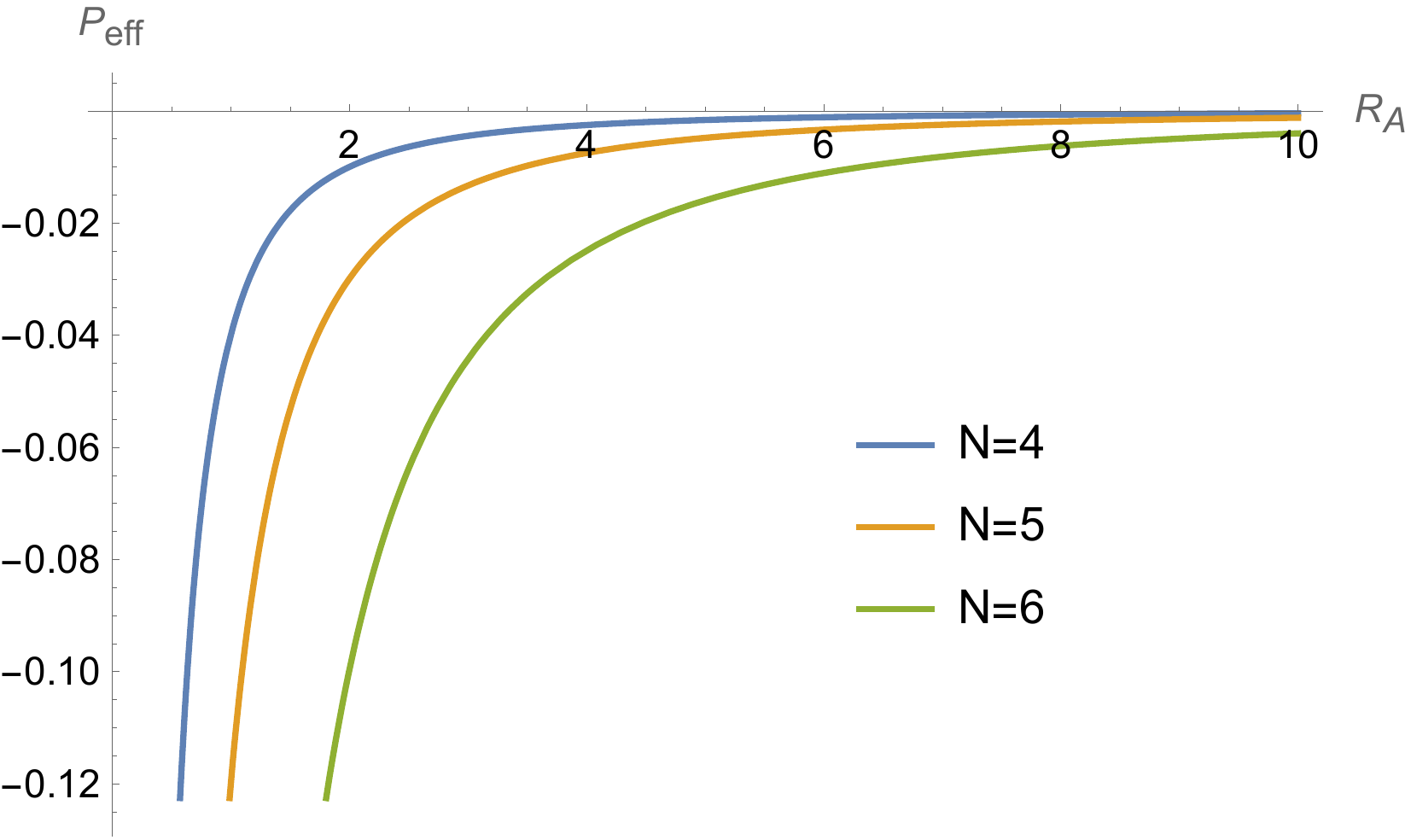}
\caption{The effective pressure of the FRW universe in Einstein gravity with $N=4,5,6$.}
\end{figure}

One can also get the `Smarr' relation of the FRW universe
\begin{alignat}{1}
E=-\frac{N-1}{N-3}P_{eff}V,
\end{alignat}
and the equation of state
\begin{alignat}{1}
P_{eff}=-\frac{N-3}{N-1}\rho,
\end{alignat}
i.e. $\omega=(3-N)/(N-1)$.

One can also define the effective enthalpy as
\begin{alignat}{1}
H:=E+P_{eff}V=-\frac{2}{N-3}P_{eff}V=\frac{N-2}{\kappa_N(N-1)}\mathcal{A}_{N-2}R^{N-3},
\end{alignat}
which satisfies
\begin{alignat}{1}
\od H=V\od P_{eff}.
\end{alignat}

If one further equips the FRW universe with a temperature $T$,
one can get the Helmholtz free energy and Gibbs free energy
\begin{alignat}{1}
F:=&E-T S_0=\frac{(N-2)}{2\kappa_N}\mathcal{A}_{N-2}R_A^{N-3}-T S_0,
\\
G:=&E+P_{eff}V-T S_0=\frac{(N-2)}{\kappa_N(N-1)}\mathcal{A}_{N-2} R_A^{N-3}-T S_0,
\end{alignat}
which can also be expressed in the `Smarr' form
\begin{alignat}{1}
F=-\frac{N-1}{N-3}P_{eff}V-T S_0, \quad G=-\frac{2}{N-3}P_{eff}V-T S_0,
\end{alignat}
and satisfy
\begin{alignat}{1}
\od F=-S_0\od T-P_{eff}\od V, \quad \od G=-S_0\od T+V\od P_{eff}.
\end{alignat}

\section{Effective Pressure of the FRW Universe in Gauss-Bonnet Gravity}

Gauss-Bonnet gravity is one of the most common and simple modified theories of gravity.
It has high order derivative terms of the curvature that combined in a certain way, i.e. the Gauss-Bonnet term,
\begin{alignat}{1}
R_{GB}=R^2-4R_{\mu\nu}R^{\mu\nu}+R_{\mu\nu\rho\sigma}R^{\mu\nu\rho\sigma},
\end{alignat}
which only has dynamical effect in $N\geq 5$.
The action of the Gauss-Bonnet gravity is written as
\begin{alignat}{1}
S=\frac{1}{16\pi G}\int\od^N x\sqrt{-g}(R+\alpha R_{GB})+S_m,
\end{alignat}
where $\alpha$ is the coupling constant and $S_m$ stands for the action of matter such as a perfect fluid.

The Friedmann's equation in Gauss-Bonnet gravity is\cite{Akbar:2006kj}
\begin{alignat}{1}
H^2+\frac{k}{a^2}+\tilde{\alpha}\left(H^2+\frac{k}{a^2}\right)^2=\frac{2\kappa_N}{(N-1)(N-2)}\rho,
\end{alignat}
where $\tilde{\alpha}=(N-3)(N-4)\alpha$, so it has no difference with Einstein gravity in $N=3,4$.
Using the apparent horizon (\ref{AH}), one can get the energy density
\begin{alignat}{1}
\rho=\frac{(N-1)(N-2)}{2\kappa_N R_A^2}\left(1+\frac{\tilde{\alpha}}{R_A^2}\right), \label{edgb}
\end{alignat}
and from the continuity equation (\ref{continue}), one can get the pressure
\begin{alignat}{1}
p=\frac{(N-2)\dot{R}_A}{\kappa_N HR_A^3}\left(1+\frac{2\tilde{\alpha}}{R_A^2}\right)
-\frac{(N-1)(N-2)}{2\kappa_N R_A^2}\left(1+\frac{\tilde{\alpha}}{R_A^2}\right).
\end{alignat}
We also give the expression of the work density
\begin{alignat}{1}
W=\frac{1}{2}(\rho-p)=\frac{(N-1)(N-2)}{2\kappa_N R_A^2}\left(1+\frac{\tilde{\alpha}}{R_A^2}\right)
-\frac{(N-2)\dot{R}_A}{2\kappa_N HR_A^3}\left(1+\frac{2\tilde{\alpha}}{R_A^2}\right).
\end{alignat}
From the energy density (\ref{edgb}), one can get the effective energy for the FRW universe in Gauss-Bonnet gravity
\begin{alignat}{1}
E=\rho V=\frac{(N-2)}{2\kappa_N}\mathcal{A}_{N-2} R_A^{N-3}\left(1+\frac{\tilde{\alpha}}{R_A^{2}}\right).
\end{alignat}
Finally, the effective pressure is obtained
\begin{alignat}{1}
P_{eff}=&-\frac{\od E}{\od V}=-\frac{(N-2)}{2\kappa_N R_A^2}\left[N-3+(N-5)\frac{\tilde{\alpha}}{R_A^{2}}\right]
\nonumber \\
=&-\frac{(N-2)(N-3)}{2\kappa_N R_A^2}-(N-2)(N-3)(N-4)(N-5)\frac{\alpha}{2\kappa_N R_A^4}, \label{PR}
\end{alignat}
which is different with the effective pressure in Einstein gravity if $N\geq 6$. It also coincide with $p$ or $-W$ if $\dot{R}_A=HR_A$ or $2HR_A$,
see Appendix. One can see that if $N\geq6$ and $\alpha>0$, the effective pressure is always negative; if $N\geq6$ and $\alpha<0$, we have $P_{eff}=0$ at $R_{A}=\sqrt{-(N-4)(N-5)\alpha}$
and a minimum $P_{min}=\frac{(N-2)(N-3)}{8\kappa_N(N-4)(N-5)\alpha}$ at $R_{A}=\sqrt{-2(N-4)(N-5)\alpha}$.
For the $\alpha<0$ case, one can define dimensionless effective pressure and apparent horizon radius
\begin{alignat}{1}
\tilde{P}_{eff}=-\alpha P_{eff}, \quad \tilde{R}_{A}=\frac{R_A}{\sqrt{-\alpha}},
\end{alignat}
and write (\ref{PR}) as
\newpage
\begin{alignat}{1}
\tilde{P}_{eff}=-\frac{(N-2)(N-3)}{2\kappa_N \tilde{R}_A^2}+\frac{(N-2)(N-3)(N-4)(N-5)}{2\kappa_N \tilde{R}_A^4}.
\end{alignat}
In the following FIG. 2, we draw the effective pressure for $N=6,7,8$.
\begin{figure}[h]
\includegraphics[height=5cm]{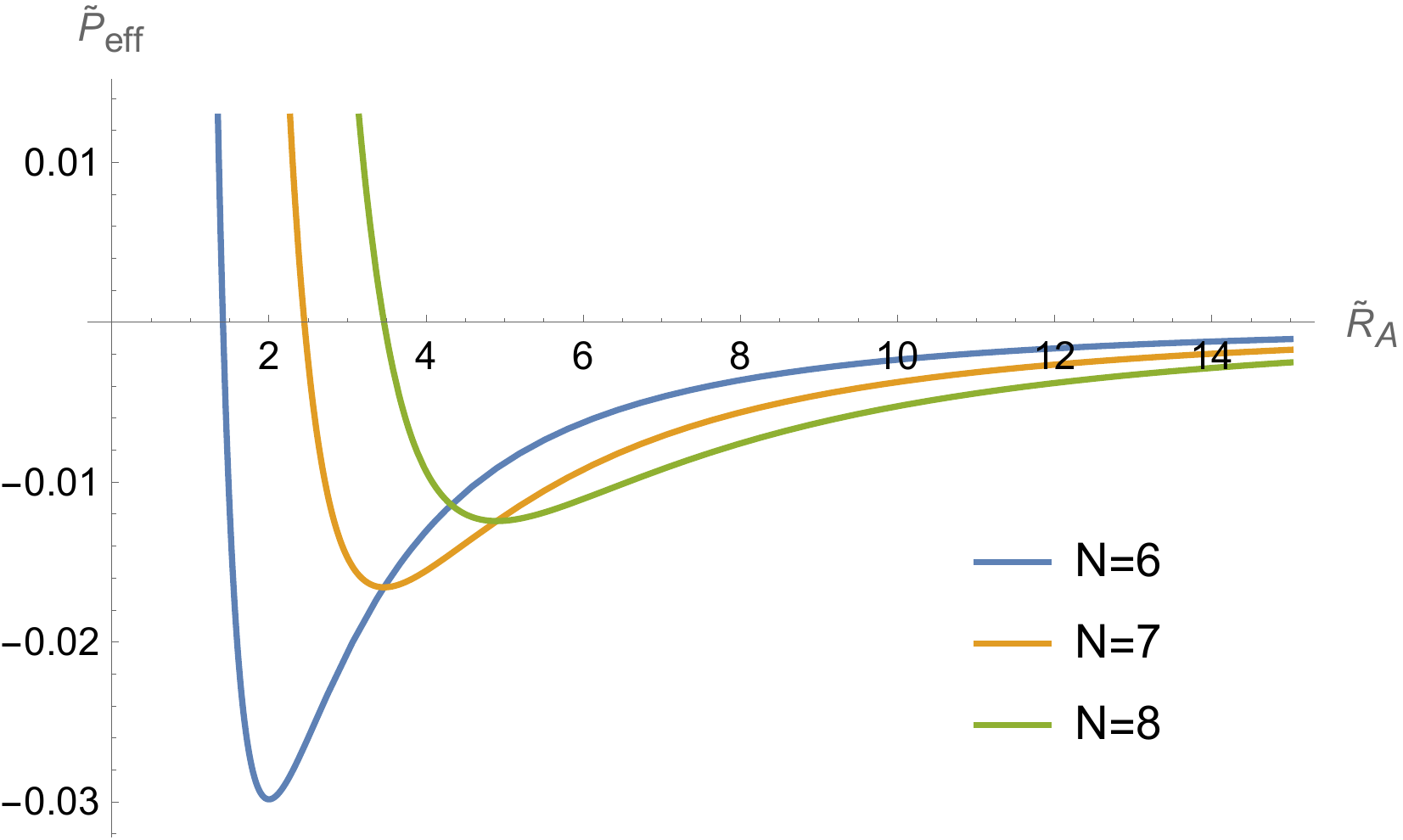}
\caption{The effective pressure of the FRW universe in Gauss-Bonnet gravity with $N=6,7,8$.}
\end{figure}

From the figure, we can see that the effective pressure behaves like the potential between molecules. Inspired by this figure, we also find that the minimum point of the curve with $N$ lies at the curve with $N-1$.

We can also define the effective enthalpy as
\begin{alignat}{1}
H:=E+P_{eff}V=\frac{(N-2)\mathcal{A}_{N-2} R_A^{N-3}}{\kappa_N(N-1)}\left(1+\frac{2\tilde{\alpha}}{R_A^2}\right),
\end{alignat}
which satisfies
\begin{alignat}{1}
\od H=V\od P_{eff}.
\end{alignat}
In this case, we also equip the FRW universe with a temperature to get the Helmholtz free energy and Gibbs free energy
\begin{alignat}{1}
F:=&E-T S_0=\frac{(N-2)}{2\kappa_N}\mathcal{A}_{N-2} R_A^{N-3}\left(1+\frac{\tilde{\alpha}}{R_A^{2}}\right)-T S_0,
\\
G:=&E+P_{eff}V-T S_0=\frac{(N-2)\mathcal{A}_{N-2} R_A^{N-3}}{\kappa_N(N-1)}\left(1+\frac{2\tilde{\alpha}}{R_A^2}\right)-T S_0,
\end{alignat}
which also satisfy
\begin{alignat}{1}
\od F=-S_0\od T-P_{eff}\od V, \quad \od G=-S_0\od T+V\od P_{eff}.
\end{alignat}

\section{Effective Pressure of the FRW Universe in Lovelock Gravity}

Lovelock gravity is a generalization of Einstein gravity and Gauss-Bonnet gravity.
The action in Lovelock gravity can be written as
\begin{alignat}{1}
S=\frac{1}{16\pi G}\int\od^{N}x\sqrt{-g}\sum_{i=0}^{m}c_i L_i+S_m,
\end{alignat}
where $c_i$ is an arbitrary constant, $m\leq[(N-1)/2]$, $S_m$ still stands for the action of matter,
and $L_i$ is the Euler density of a $2i$-dimensional manifold\cite{Cai:2006rs}
\begin{alignat}{1}
L_i=\frac{1}{2^i}\delta^{a_1b_1...a_ib_i}_{c_1d_1...c_id_i}R^{c_1d_1}_{~~~~a_1b_1}...R^{c_id_i}_{~~~~a_ib_i}.
\end{alignat}
The cosmological constant corresponds to $i=0$, the Einstein-Hilbert term corresponds to $i=1$, and the Gauss-Bonnet term corresponds to $i=2$.
From the action, one can get the field equations
\begin{alignat}{1}
\sum_{i=0}^{m}\frac{c_i}{2^{i+1}}\delta^{aa_1b_1...a_ib_i}_{bc_1d_1...c_id_i}R^{c_1d_1}_{~~~~a_1b_1}...R^{c_id_i}_{~~~~a_ib_i}=\kappa_N T^a_b,
\end{alignat}
where $T^a_b$ can still adopt the form of a perfect fluid (\ref{pf}).

From the field equations, one can get the Friedmann's equation
\begin{alignat}{1}
\sum_{i=1}^{m}\hat{c}_i\left(H^2+\frac{k}{a^2}\right)^{i}=\frac{2\kappa_N}{(N-1)(N-2)}\rho, 
\end{alignat}
where
\begin{alignat}{1}
\hat{c}_i=\frac{(N-1)!}{(N-1-2i)!}c_i,
\end{alignat}
combined with (\ref{AH}), one can get
\begin{alignat}{1}
\sum_{i=1}^{m}\frac{\hat{c}_i}{R_A^{2i}}=\frac{2\kappa_N}{(N-1)(N-2)}\rho. \label{felove}
\end{alignat}
From the continuity equation (\ref{continue}), we get the pressure of the perfect fluid
\begin{alignat}{1}
p=\frac{N-2}{\kappa_N H}\sum_{i=1}^{m}i\hat{c}_i\left(\frac{1}{R_A}\right)^{2i+1}\dot{R}_A
-\frac{(N-1)(N-2)}{2\kappa_N}\sum_{i=1}^{m}\hat{c}_i\left(\frac{1}{R_A}\right)^{2i}.
\end{alignat}
We also give the expression of the work density
\begin{alignat}{1}
W=\frac{1}{2}(\rho-p)=\frac{(N-1)(N-2)}{2\kappa_N}\sum_{i=1}^{m}\hat{c}_i\left(\frac{1}{R_A}\right)^{2i}
-\frac{N-2}{2\kappa_N H}\sum_{i=1}^{m}i\hat{c}_i\left(\frac{1}{R_A}\right)^{2i+1}\dot{R}_A.
\end{alignat}

From the equation (\ref{felove}), we get the effective energy
\begin{alignat}{1}
E=\rho V=\frac{(N-2)\mathcal{A}_{N-2}}{2\kappa_N}\sum_{i=1}^{m}\hat{c}_i R_A^{N-1-2i},
\end{alignat}
so the effective pressure is
\begin{alignat}{1}
P_{eff}=-\frac{\od E}{\od V}=-\frac{(N-2)}{2\kappa_N}\sum_{i=1}^{m}\frac{\hat{c}_i(N-1-2i)}{R_A^{2i}},
\end{alignat}
which still coincides with $p$ or $-W$ if $\dot{R}_A=HR_A$ or $2HR_A$ respectively, see Appendix.\footnote{For all three cases, i.e. Einstein gravity, 
Gauss-Bonnet gravity, Lovelock gravity, the two conditions are the same, which can be derived in a general way, see Appendix.}
The effective pressure consists of many terms, so it may have multiple zero-points and extreme points.

The effective enthalpy is
\begin{alignat}{1}
H:=E+P_{eff}V=\frac{(N-2)\mathcal{A}_{N-2}}{\kappa_N(N-1)}\sum_{i=1}^{m}i\hat{c}_i R_A^{N-1-2i},
\end{alignat}
which satisfies
\begin{alignat}{1}
\od H=V\od P_{eff}.
\end{alignat}

In this case, we also equip the FRW universe with a temperature to get the Helmholtz free energy and Gibbs free energy
\begin{alignat}{1}
F:=&E-T S_0=\frac{(N-2)\mathcal{A}_{N-2}}{2\kappa_N}\sum_{i=1}^{m}\hat{c}_i R_A^{N-1-2i}-T S_0,
\\
G:=&E+P_{eff}V-T S_0=\frac{(N-2)\mathcal{A}_{N-2}}{\kappa_N(N-1)}\sum_{i=1}^{m}i\hat{c}_i R_A^{N-1-2i}-T S_0,
\end{alignat}
which also satisfy
\begin{alignat}{1}
\od F=-S_0\od T-P_{eff}\od V, \quad \od G=-S_0\od T+V\od P_{eff}.
\end{alignat}

\section{Conclusion and Discussion}

In this paper, we find a new (effective) definition of pressure for the FRW universe. It is defined by
$P_{eff}=-\od E/\od V$, here $E=\rho V$ is the effective energy, $\rho$ is the energy density, and $V$ is the volume of the FRW universe inside
the apparent horizon. From the Friedmann's equations in Einstein gravity, Gauss-Bonnet gravity, and Lovelock gravity,
we get the effective pressure, which is equivalent to the `ordinary' pressure $p$ when $\dot{R}_A=H R_A$ regardless of the choice of gravitational theories.

The effective pressure in Einstein gravity is always negative and increases with the radius of the apparent horizon $R_A$.
The effective pressure in Gauss-Bonnet gravity is different from the one in Einstein gravity if $N\geq 6$. 
If $\alpha>0$, the effective pressure is always negative; if $\alpha<0$, the effective pressure is positive when $R_A<\sqrt{-(N-4)(N-5)\alpha}$ or negative when $R_A>\sqrt{-(N-4)(N-5)\alpha}$ and has a minimum $P_{min}=\frac{(N-2)(N-3)}{8\kappa_N(N-4)(N-5)\alpha}$ at $R_A=\sqrt{-2(N-4)(N-5)\alpha}$. Interestingly, the relation between the effective pressure and the radius of the apparent horizon is very similar to the relation between the potential and distance of molecules. More interestingly, the minimum point of the effective pressure in $N$-dimension coincides with the effective pressure in $(N-1)$-dimension with the same apparent horizon. The effective pressure in Lovelock gravity consists of many terms, which may have multiple zero-points and extreme points. We also give the expressions of enthalpy and free energy based on the effective pressure of the FRW universe in these three theories of gravity.

The results show that both gravity and dimension can affect the effective pressure of the FRW universe, and the effective pressure
in different dimensions also has an interesting relation. The effective pressure is a function of the radius of the apparent horizon or
the volume of the FRW universe inside the apparent horizon. Naturally, the effective pressure can be introduced in other dynamical
spacetimes such as dynamical black holes. For stationary black holes such as Schwarzschild black hole, it may also be used if we
treat them as extreme cases of the dynamical black holes that change very slowly.

\section*{Acknowledgments}

I thank for the discussions with Yi-Hao Yin, Yang Liu, and Alexey Pronin. This work is supported by National Natural Science Foundation of China (NSFC) 
under No.12465011 and East China University of Technology (ECUT) under DHBK2023002.

\appendix

\section{A General Derivation of the Conditions of $P_{eff}=p$ and $P_{eff}=-W$ }

In all the three kinds of gravity, i.e. Einstein gravity, Gauss-Bonnet gravity, Lovelock gravity, we have
\begin{alignat}{1}
P_{eff}=-\frac{\od E}{\od V}=-\frac{\od (\rho V)}{\od V}=-\frac{\od\rho}{\od V}V-\rho
=-V\frac{\od\rho}{\od t}/\frac{\od V}{\od t}-\rho,
\end{alignat}
combined with $V=\mathcal{A}_{N-2} R_A^{N-1}/(N-1)$ and the continuity condition $\dot{\rho}+(N-1) H(\rho+p)=0$, we can get
\begin{alignat}{1}
P_{eff}=\frac{HR_A}{\dot{R}_A}(\rho+p)-\rho.
\end{alignat}
Therefore, it can be easily seen that if $\dot{R}_A=HR_A$, one has $P_{eff}=p$,
and if $\dot{R}_A=2HR_A$, one has $P_{eff}=(p-\rho)/2=-W$.

\end{document}